\documentclass[aps,prb,twocolumn,groupedaddress,showpacs, amsmath,amssymb,]{revtex4-1}
\usepackage{graphicx}
\usepackage{amssymb}
\usepackage{amsmath}
\usepackage{dcolumn}
\usepackage{xcolor}
\usepackage{multirow}
\usepackage{float}

\begin{document}


\title{Impact of Rare-earth site Substitution on the Structural, Magnetic and Thermal Properties   of  Ce$_{1-x}$Eu$_x$CrO$_3$ Orthochromite Solid-solutions}

\author{M. Taheri$^\dagger $}
\affiliation{ Department of Physics, Brock University, St Catharines, ON, L2S 3A1, Canada}

\author{F. S. Razavi}
\affiliation{ Department of Physics, Brock University, St Catharines, ON, L2S 3A1, Canada}
\author{Z. Yamani}
\affiliation{ Chalk River Nuclear Laboratories, Chalk River, ON, K0J 1J0, Canada}
\author{R. Flacau$^\ddagger$}
\affiliation{ Chalk River Nuclear Laboratories, Chalk River, ON, K0J 1J0, Canada}

\author{C. Ritter}
\affiliation{Institut Laue-Langevin, F-38042 Grenoble, France}

\author{S. Bette}
\affiliation{ Max Planck Institute for Solid State Research, Stuttgart, Germany}

\author{R. K. Kremer}
\affiliation{ Max Planck Institute for Solid State Research, Stuttgart, Germany}

\begin{abstract}

The role of  slight changes of the chemical composition on antiferromagnetic ordering of Cr in rare-earth orthochoromites was investigated on a series of ceramic solid-solutions Ce$_{1-x}$Eu${_x}$O$_3$ where x varied from 0 to 1. Gradual replacement of Ce with Eu  reduces the cell volume and acts equivalently to applying external pressure. Full replacement of Ce by Eu, on the other hand, reduces the N\'{e}el temperature from 260 K for CeCrO$_3$ to 178 K for  EuCrO$_3$ as established by magnetization, heat capacity and neutron powder diffraction measurements.
High resolution x-ray powder diffraction measurements on   Ce$_{1-x}$Eu${_x}$O$_3$ and neutron powder diffraction studies  on CeCrO$_3$ enable to correlate the magnetic properties of the Cr magnetic subsystem with the size of the lattice and minute changes of the  bonding and torsion angles within and between the CrO$_6$ octahedra. We find that the sizes and the shapes of the CrO$6$ octahedra remain essentially unchanged as the size of the rare-earth cations is reduced whereas decreasing Cr - O - Cr bonding angles and increasing inclination of neighboring octahedra enable  to compensate for the decreasing lattice size.
\end{abstract}

 \pacs{75.25.-j, 75.47.Lx, 65.40.Ba, 65.40.De, 75.50.Ee}

 \maketitle

\section{Introduction}

\begin{figure}[htp]
	\centering
   \includegraphics[height=7cm]{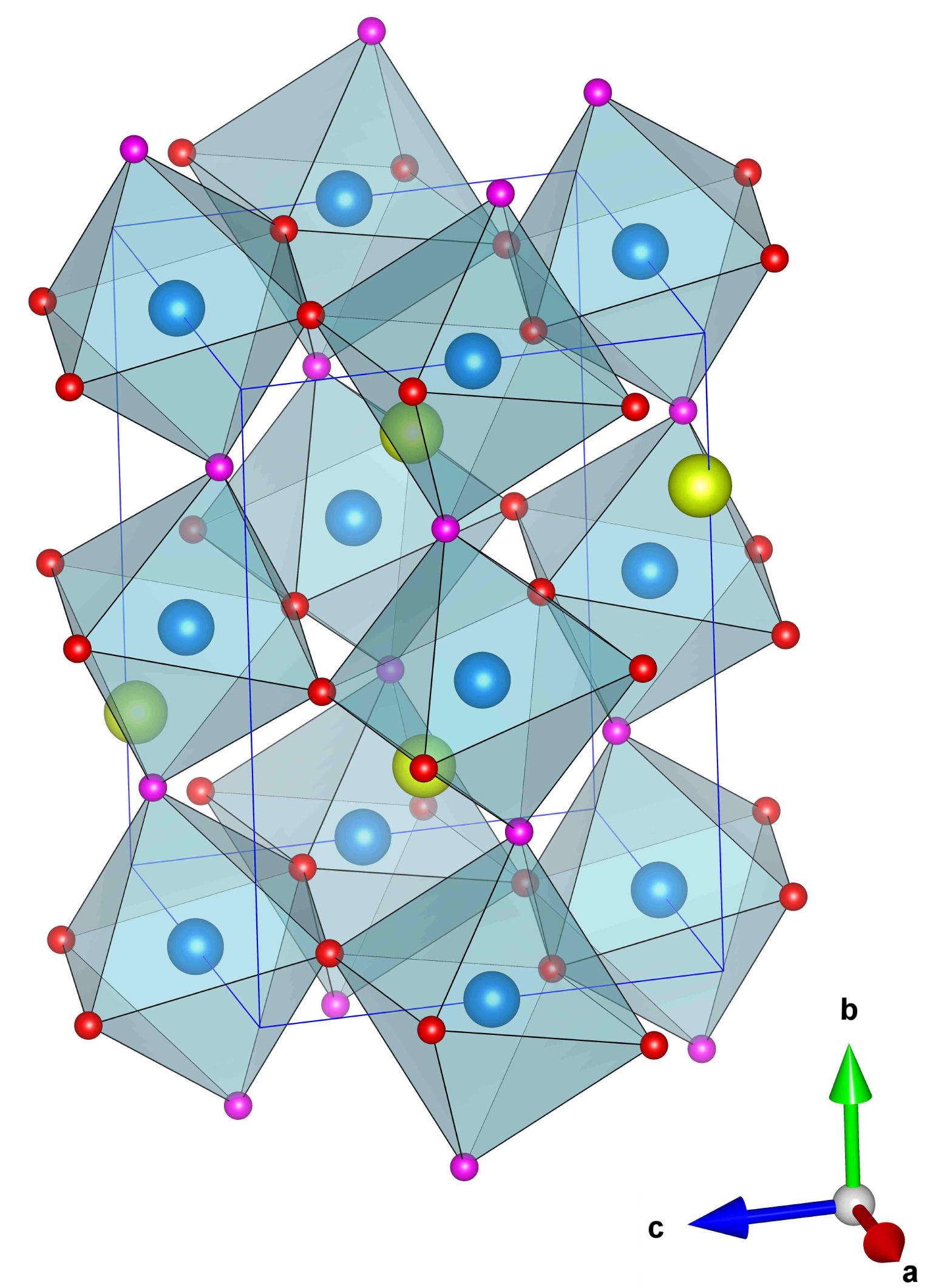}
	\caption{(color online)  Crystal structure of (Ce,Eu)CrO$_3$ in space group $Pnma$ (no. 62). Yellow/green and blue spheres represent cerium/europium and chromium atoms, respectively.  The O2 atoms connecting the CrO$_6$ octahedra  in the $a$ - $c$ plane are given by red spheres and the O1 atoms connecting along $b$ are represented by magenta spheres. A unit cell is outlined.}
	\label{Fig1}
\end{figure}

Rare-earth orthochromites which crystallize  with the orthorhombic GdFeO$_3$ structure-type (Figure \ref{Fig1}) have attracted special attention, because they may
exhibit  ferroelectric polarization either induced by an external magnetic field  or spontaneous polarization due to internal magnetic fields induced by long-range magnetic ordering.\cite{Sahu2007,Rajeswaran2012,Raveau2014,Saha2014,Meher2014,McDannald2015}
For example, magneto-electric effects as well as magnetic and electric field induced switching of the dielectric polarization have been detected in SmCrO$_3$, GdCrO$_3$, and ErCrO$_3$.\cite{Rajeswaran2012}  Still controversial is the relevance of the interaction between the rare-earth magnetism and the weak ferromagnetism due to spin canting in the Cr subsystem. For example, polarization in ErCrO$_3$ vanishes at a spin re-orientation transition at $\sim$22 K below which spin canting of the Cr moments also disappears emphasizing  the importance of weak ferromagnetism and the interaction with the rare-earth subsystem for polar ordering.\cite{Rajeswaran2012}
For LuCrO$_3$, on the other hand,  Preethi Meher \textit{et al.} observed qualitatively similar ferroelectric-like characteristics as in ErCrO$_3$, though weaker in magnitude,  revealing that orthochromites with a diamagnetic rare-earth constituent can also develop a polar state questioning the necessity of  rare-earth magnetism  for establishing multiferroicity in the orthochromites.\cite{Meher2014}
At present, there are also diverging conclusions  as to whether the multiferroic state as evidenced by polarization and pyroelectric currents occurs below the N\'eel temperature \cite{Rajeswaran2012} or whether it is already formed  at higher temperatures.\cite{Lal1989,Sahu2007} Ghosh \textit{et al.} have observed that in SmCrO$_3$ and HoCrO$_3$  polar order already develops in the paramagnetic state and they attributed this observation to a structural transition from the centrosymmetric orthorhombic space group $Pbnm$  to the non-centrosymmetric space group $Pna$2$_{\rm 1}$.\cite{GhoshSmCrO3,GhoshHoCrO3}  Kumar \textit{et al.}\cite{Kumar2016} recently, suggested that an additional driving force for observation of ferroelectricity in rare-earth orthochromites is the short range exchange interactions.\cite{Wan2016}

The range of the magnetic ordering temperatures of the RCrO$_3$  opens up potential  applications for these compounds.
Compared to what has been observed, for example, in the orthomanganites family, RMnO$_3$, (e.g. $T_{\rm N} \sim$ 27 K in TbMnO$_3$) which comprises some of the most prominent multiferroic systems, the rare-earth orthochromites exhibit ordering temperatures close to room temperature. Also, since the coupling between the 4$f$  and the 3$d$ electrons in the half filled $t_{\rm 2g}$ orbitals of the Cr$^{3+}$ cations can be large, orthochromites  may have some advantages compared to orthoferrites. Although the latter exhibit higher magnetic ordering temperatures, Cr$^{3+}$ lacks  fourth order crystal field splitting terms acting on the $S$ =$\frac{3}{2}$ spin multiplet  (cf. Ref. \onlinecite{Hornreich1978}). This might allow higher complexity of spin canting and spin reorientation phenomena
in the magnetic structures of the rare-earth orthochromites.

The structural and magnetic properties of the rare-earth orthochromites, RCrO$_3$,  have been studied in some detail before.\cite{Zhou2010,Gonjal2013,Shukla2009,MaryamPRB}
By using high-resolution neutron powder diffraction, Zhou \textit{et al.} determined the crystal structures  including the oxygen atom positions  with great precision.\cite{Zhou2010} They found that the cooperative rotation and tilting of the CrO$_6$ octahedra essentially tunes the $t_{\rm 2g}$ - $e_{\rm g}$ overlap and the Cr - O - Cr superexchange which can explain the sizeable decrease of the N\'{e}el temperature when substituting smaller rare-earth cations.
The  N\'{e}el temperature, $T_{\rm N}$, of the Cr substructure which is at about 300 K for LaCrO$_3$
 decreases monotonically as the rare-earth element is replaced by a heavier one  with $T_{\rm N}$ for LuCrO$_3$ of about 100 K.\cite{Gonjal2013}
CeCrO$_3$, studied  by Shukla \textit{et al.}, exhibits a N\'{e}el temperature of $\sim$260 K  close to that  of LaCrO$_3$.\cite{Shukla2009}
EuCrO$_3$ orders at $\sim$178 K with a slightly canted antiferromagnetic structure.\cite{MaryamPRB}

The marked decrease of the N\'{e}el temperature of the Cr substructure in the series from the light to the heavy rare-earth elements with the simultaneous reduction of the unit cell volume  already  were suggestive that  the Cr - Cr distances alone cannot be the determining factor for the Cr magnetic ordering in the RCrO$_3$.
Rather, as concluded by Zhou \textit{et al.} bonding and  tilting angles of the CrO$_6$ octahedra  change and the crystal structure  becomes more distorted, reducing Cr - O - Cr electron transfer and in turn superexchange between the Cr cations.\cite{Zhou2010}
Bhadram \textit{et al.} have investigated the variation of the room temperature lattice parameters and Raman frequencies of RCrO$_3$ (R = Eu, Gd, Tb, Lu) with pressure and found that the octahedral tilts decrease linearly with increasing R radii.\cite{Bhadram2014} Compressibility increases for small R ions whereas the structures of rare-earth orthochromites with larger R ions are stiffer. They  concluded that for the N\'{e}el temperature the Cr - O bond lengths play a dominant role over the octahedral tilts.\cite{Bhadram2014}
Zhao \textit{et al.} using first-principles calculations found opposite dependence of the N\'{e}el temperature on hydrostatic or chemical pressure.\cite{Zhao2013} Whereas, hydrostatic pressure increases the N\'{e}el temperature chemical pressure e.g. induced by replacing larger rare-earth cations by smaller ones lowers the N\'{e}el temperature, in qualitative agreement with the N\'{e}el temperatures observed across the series of the rare-earth orthochromites.

For CeCrO$_3$ ($T_{\rm N} \sim$260 K) Shukla \textit{et al.} found a G-type magnetic ordering of the Cr substructure
and  also proposed  ordering of the Ce moments  below $\sim$100 K and a spin re-orientation below $\sim$15 K.  However, this conclusion was based on  very faint additional magnetic scattering peak in the neutron powder diffraction patterns at temperatures  well below $\sim$100 K.\cite{Shukla2009} The refined Ce moments amounted to $\sim$0.7 $\mu_{\rm B}$ and pointed antiparallel to the Cr moments.
In preceding publications, we have investigated in detail the magnetic properties of  EuCrO$_3$ and CeCrO$_3$.\cite{MaryamPRB,MaryamJAP} Below 178 K, the Cr substructure in EuCrO$_3$ orders with a $G_x$-type antiferromagnetic structure with an ordered moment of $\sim$2.4 $\mu_{\rm B}$, consistent with
the $S$ = $\frac{3}{2}$ spin-only ground state. A weak canted ferromagnetic moment pointing along the $c$-axis ($Pbnm$ notation) could be concluded from the neutron diffraction investigation. It amounts to $\sim$0.1 $\mu_{\rm B}$, consistent with single crystal magnetization data.\cite{Tsushima1969} Evidence for
ordering of the Eu moments has neither been found in the neutron powder diffraction data nor in the heat capacity or magnetization data. This observation can be understood as being due to the $J$ = 0 Hund's rule ground state and the van Vleck-type temperature dependent moment of the Eu$^{3+}$ cations.\cite{MaryamPRB}
In contrast to the absence of conclusive evidence for R moment ordering in CeCrO$_3$ and EuCrO$_3$, there is some unusual behavior of the magnetic susceptibilities at low temperature.\cite{Shukla2009,MaryamJAP} Especially, the field-cooled (FC) susceptibility of CeCrO$_3$ showed a compensation point at $\sim$50 K with  negative magnetization below. Irrespective of the measuring field, zero-field cooled (ZFC) and FC magnetic susceptibilities intersected at $\sim$100 K with the FC susceptibility being smaller than the ZFC susceptibility. This complex behavior could be due to  inhomogeneity in the small sample grains (typically $<$ 100 nm, "core-shell" behavior) or be due to some complex compensation of R and Cr moments induced by external fields.\cite{MaryamJAP}

We explore in this paper the changes of the magnetic and structural properties induced by the replacement of a larger rare-earth cation Ce$^{3+}$ by the smaller one Eu$^{3+}$ with ionic radii of 1.283 \AA  and  1.144 \AA, respectively.\cite{Shannon1976}
Especially, we study in detail the changes of the crystal structure and magnetic properties in small steps with a series of solid-solutions of CeCrO$_3$ and EuCrO$_3$,
the latter having  unit cell volumes of 230.8 \AA$^3$ and 224.8 \AA$^3$, respectively. Gradual replacement of Ce with Eu leads to an increase of the cell distortion factor\cite{Sasaki1983}  by $\sim$ 50\% similar to the effect of external hydrostatic pressure of $\sim$10 GPa or more.\cite{Bhadram2014}.   On the other hand, EuCrO$_3$ has a markedly lower N\'{e}el temperature than CeCrO$_3$ suggesting that substituting Ce by the smaller Eu cations i.e. chemical pressure will act differently from hydrostatic pressure, supporting the first-principles results obtained by Zhao \textit{et al.}.\cite{Zhao2013}

The paper is organized in two parts, followed by a summary: We first describe structural properties of the Ce$_{1-x}$Eu$_x$CrO$_3$ phases across the entire solid-solution range 0 $\leq x \leq $ 1.
This set of solid-solutions allows to tune the mean ionic radius of the rare-earth cations between those of Ce and Eu, thus covering continuously a range which was accessed by
high-resolution neutron powder diffraction for NdCrO$_3$ and PrCrO$_3$ only. Samples containing Sm, Eu, and Gd orthochromites were not included in this series.\cite{Zhou2010} In the second part, we present the results of our temperature, field and thermal history  dependent of dc and ac magnetic susceptibilities and thermal properties and report first neutron  powder diffraction investigations on the solid solutions Ce$_{1-x}$Eu$_x$CrO$_3$. Finally, we discuss the results and summarize.

\section{Experiment}
Powder samples of Ce$_{1-x}$Eu$_x$CrO$_3$ were synthesized by the solution combustion method starting from equimolar solutions of high purity cerium and europium nitrate, chromium nitrate and glycine as described in detail elsewhere.\cite{MaryamJAP,MaryamPhD} The products were characterized with respect to  phase purity by x-ray powder diffraction (XRPD) at room temperature using  Mo$K_{\alpha 1}$ radiation (2$\theta_{max}\sim$110$^{\rm o}$, STOE triple DECTRIS Multi-MYTHEN silicon strip detector).  Neutron powder diffraction (NPD) patterns of CeCrO$_3$ were collected at room temperature on the high-resolution two-axis diffractometer D2B installed at the Institut Laue-Langevin (ILL, Grenoble) using neutrons of 1.594 {\AA} wavelength. The polycrystalline sample of $\sim$9 g was contained in a thin-walled vanadium container of 8 mm outer diameter. Rietveld refinements by varying lattice parameters and atom positions were performed in the orthorhombic spacegroup $Pnma$ (no. 62) using the $FullProf$ or the TOPAS software \cite{Rietveld,FullProf,Topas}  assuming a Thompson-Cox-Hastings pseudo-Voigt peak profile ($FullProf$ NPR = 7).  Isotropic and anisotropic displacement parameters were tested in case of the x-ray and neutron diffraction patterns, respectively.  Since marked anisotropies could not be seen we report isotropic displacement parameters in the following. The background was modeled by a higher-order Chebychev polynomial.  The refinements  typically converged to Bragg- and R$f$-reliability factors of $\sim$ 2\% or less and $\chi^2$-values better than 1 were achieved for the x-ray patterns. For the neutron diffraction pattern of CeCrO$_3$, the reliability factors were somewhat larger (see Table \ref{Table2}).
Medium resolution high intensity neutron scattering was performed on the N5 triple axis spectrometer  at the Chalk River Laboratories in order to follow the temperature dependence of magnetic Bragg peaks.  Initial and final neutron energies  were chosen by the pyrolytic graphite PG002 reflections. In order to reduce the high absorption cross section for thermal neutrons from the $^{151}$Eu isotope in the natural isotope composition of Eu,  the samples with a Eu content larger than 10\% were contained in flat geometry thin-walled aluminum sample holders which  were long enough to cover the full beam height, but  reduced the thickness of the sample to $\sim$ 1 mm.\cite{Ryan} dc magnetization were measured using a Magnetic Property Measurement System (Quantum Design, MPMS).  Heat capacity measurements were performed on compressed pellets in a Physical Property Measurement System (Quantum Design, PPMS). The particle size of the powders was determined with a Hitachi H-7650 transmission electron microscopy (TEM) and a Zeiss Sigma VP scanning electron microscopy (SEM) equipped with the Oxford Inca EDX.

\section{Results}
\subsection{Chemical and Structural Characterization}
\begin{figure}[htp]
	\centering
	\includegraphics[width=8.5cm]{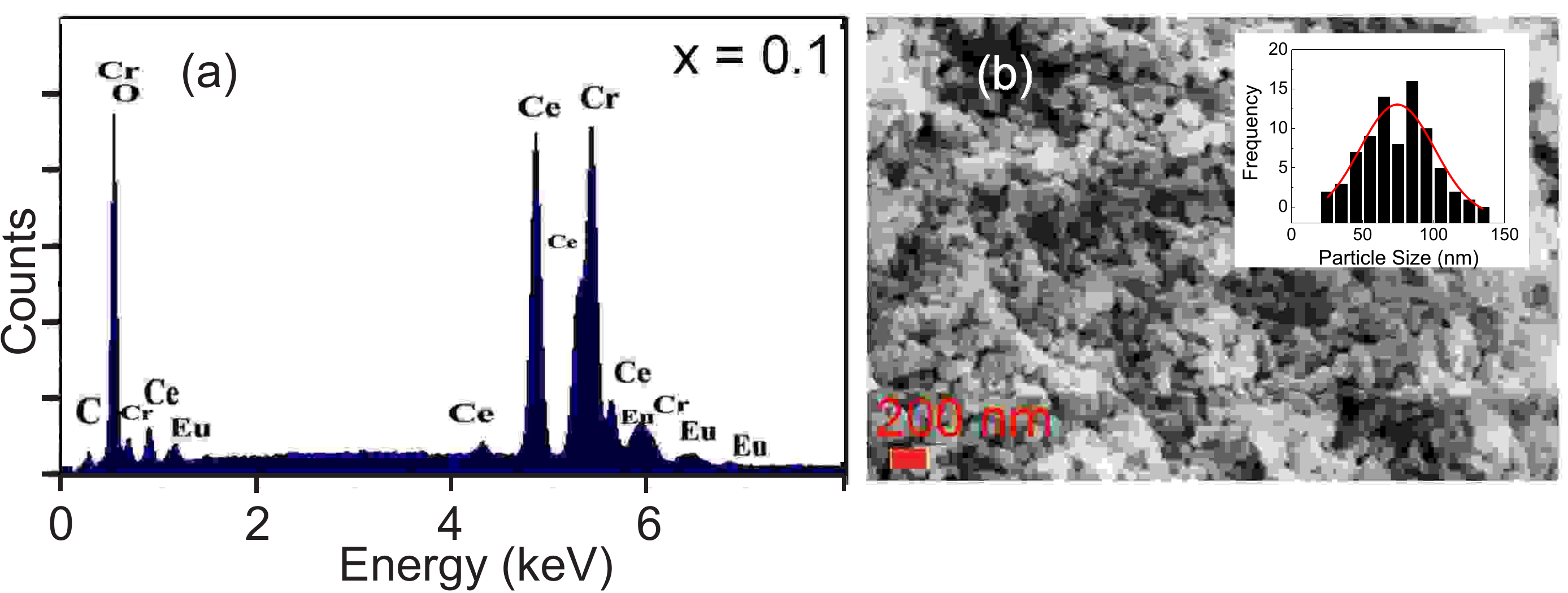}
	\caption{(color online) (a) EDX diagram of the phase Ce$_{0.9}$Eu$_{0.1}$CrO$_3$. (b) TEM image with corresponding size distribution histogram (inset) of the same sample. The red solid line is a Gaussian centered at $\sim$75 nm and a FWHM of $\sim$50 nm.}
	\label{FigSEM}
\end{figure}

Chemical and structural properties and the morphology of the boundary phases CeCrO$_3$ and EuCrO$_3$ as well as a surface analysis of the Ce$_{1-x}$Eu$_x$CrO$_3$ nano-powders have been described in detail elsewhere.\cite{MaryamJAP,MaryamPhD,MaryamPRB,MaryamXPS} Figure \ref{FigSEM} summarizes typical result of our EDX microprobe analysis and the TEM picture of the sample Ce$_{0.9}$Eu$_{0.1}$CrO$_3$. A  statistical analysis of the sample grain sizes indicated  a Gaussian distribution with a maximum at $\sim$75 nm and a FWHM of $\sim$50 nm, similar to what has been reported for pure CeCrO$_3$ and EuCrO$_3$.\cite{MaryamJAP,MaryamPhD,MaryamPRB}

\begin{figure}[htp]
	\centering
	\includegraphics[width=8.5cm]{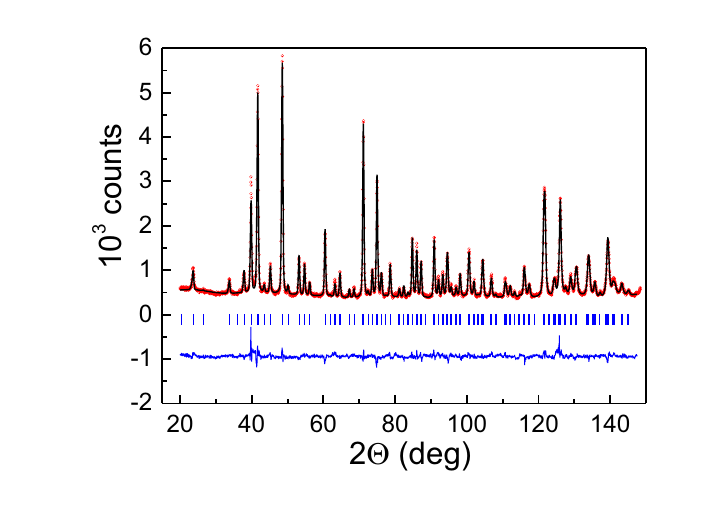}
	\caption{(color online) The NPD patterns of CeCrO$_3$  collected at room-temperature using neutrons with a wavelength of $\lambda$ =  1.594 {\AA}. Black circles represent the measured data, the red solid line is the calculated pattern of the Rietveld refinement. The blue solid line at the bottom of the graph shows the difference between the observed and calculated patterns. Vertical tics (blue) mark the position of the Bragg reflections used to simulate the refined patterns.}
	\label{NPDD2B}
\end{figure}

\begin{figure}[htp]
	\centering
	\includegraphics[width=8.5cm]{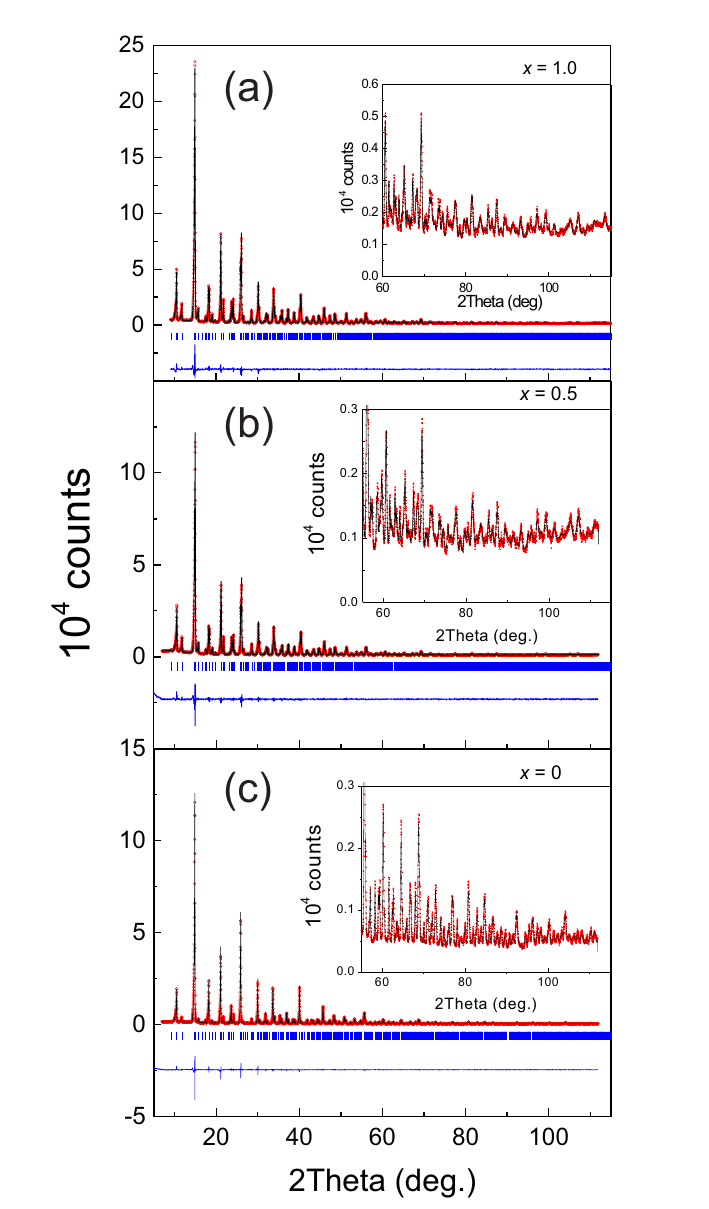}
	\caption{(color online) The XRPD patterns of Ce$_{1-x}$Eu$_x$CrO$_3$ ($x$ = 0.0, 0.5, 1.0, from bottom to top) collected at room-temperature with $\lambda$ = 0.709319{\AA} (Mo K$_{\alpha 1}$). Black circles represent the measured data, the red solid line is the result of the Rietveld refinement. The blue solid line at the bottom of the graph shows the difference between the observed and calculated patterns. Vertical tics (blue) mark angles of the Bragg reflections used to simulate the refined patterns.}
	\label{Survey}
\end{figure}

The results of our NPD on CeCrO$_3$ and the XRPD experiments on  Ce$_{1-x}$Eu$_x$CrO$_3$ together with the Rietveld refinements are shown in Figures \ref{NPDD2B} and \ref{Survey}(a - c).
The comparison of the refined structural parameters compiled in Table \ref{Table2} reveals   good agreement between parameters refined from our XRPD and NPD patterns and previously reported data.\cite{Shukla2009} Merely some atom positional parameters of the two oxygen atoms O1 (Wyckoff site 4$c$) and O2  (Wyckoff site 8$d$) differ slightly
which also explains   the small differences in the Cr - O - bonding angles. The angle $\alpha$ which reflects the O21 - Cr - O22 bonding angle in the equatorial plane  of the CrO$_6$ octahedra and the angle $\beta$ which measures the the tilt of the Cr - O1 direction with respect to the plane spanned by the O2 atoms and the Cr atoms are very close to 90$^{\rm o}$ and less different and agree well for both techniques and previously published data.\cite{Shukla2009}
The characteristic Cr - O  distances and O - Cr - O bonding angles are compiled in Table \ref{Table3}.

\begin{table}[htp]
\caption{Structural parameters and conventional reliability factors  as obtained from  Rietveld profile refinements of our XRPD and NPD patterns collected at 295 K with Mo-$K_{\alpha 1}$ radiation  and neutrons with a wavelength of  $\lambda$ = 1.549 \AA\ on CeCrO$_3$. The Rietveld refinements were performed assuming the space group $Pnma$ (no. 62). The respective site occupancies were not refined from the XRPD pattern whereas the refinement of the NPD pattern indicated full occupation as of the oxygen sites. Data by Shukla \textit{et al.}\cite{Shukla2009} are  given in the rightmost column.}
\label{Table2}
\begin{ruledtabular}
\begin{tabular}{c c  c c }
 T (K)  &   295 (XRPD) &   295 K (NPD) & 300 K (NPD)\cite{Shukla2009} \\
\hline
a  ({\AA}) & 5.4736(1) &  5.4729(2) & 5.472(1) \\
b  ({\AA})  &  7.7311(1) & 7.7304(3) & 7.733(1)  \\
c  ({\AA})  & 5.4818(1) &  5.4808(2) & 5.479(1)  \\
V  ({\AA}$^3$) & 231.97(1)& 231.88(2) & 231.84(11) \\
& & & \\
Ce  (4c) & &  & \\
x & 0.5280(1) & 0.5272(6) & 0.5294(20)\\
y & 1/4  &  1/4  & 1/4\\
z  & 0.5050(2)  &  0.5046(2) & 0.5054(9)\\
$B_{iso}$ ({\AA}$^2$)  & 0.35(1)  & 0.32(1) & 0.5\\
& & &\\
Cr (4b) &  & &  \\
x & 0 &   0 & 0 \\
y & 0 & 0  & 0 \\
z & 1/2 &  1/2 & 1/2 \\
$B_{iso}$ ({\AA}$^2$)  & 0.24(1) & 0.23(1) & 0.5 \\
& & &\\
O1 (4c) &   & &  \\
x & 0.9856(7)  & 0.9902(7) & 0.9948(15)  \\
y &  1/4 &  1/4 & 1/4 \\
z & 0.4220(15) & 0.4304(7) & 0.4260(14)   \\
$B_{iso}$ ({\AA}$^2$)  & 0.23(5)  & 0.24(1) & 0.9  \\
& & &\\
O2 (8d) &  & &   \\
x & 0.2831(9)  & 0.2834(4) & 0.2821(7)   \\
y & 0.0358(7) &  0.0391(3) & 0.0389(6)  \\
z & 7192(9) & 0.7146(5) & 0.7201(8)    \\
$B_{iso}$ ({\AA}$^2$)& 0.23(5)  & 0.29(1) & 0.9  \\
\hline
Bragg R-factor (\%) & 1.54 & 3.94 & \\
$R_f$-factor (\%) & 1.86 & 2.96 &\\
\end{tabular}
\end{ruledtabular}
\end{table}

\begin{table}[htp]
\caption{Interatomic distances and bonding angles of CeCrO$_3$ calculated from the parameters given in Table \ref{Table2}. The angles $\alpha$ and $\beta$ are defined as the
O21 - Cr - O22 bonding angle in the midsection  of the CrO$_6$ octahedra and the torsion angle $\beta$ is given by the tilt of the Cr - O1 direction with respect to a neighboring Cr - O1 bonds.}
\label{Table3}
\begin{ruledtabular}
\begin{tabular}{c c c  }
  & XRPD & NPD \\
\hline
Cr - O21 ({\AA}) & 1.964(5)  & 1.969(3)  \\
Cr - O22  ({\AA}) & 1.980(5)   & 1.986(3)   \\
Cr - O1  ({\AA}) & 1.981(2) & 1.9706(8)  \\
$\alpha$ ($^{\rm o}$) & 88.72(16) & 88.65 (11)\\
$\beta$ ($^{\rm o}$) & 89.61(5) & 89.25(4)\\
$\angle$ O1 - Cr - Cr - O1 ($^{\rm o}$) & 15.61(4) & 16.7(4) \\
$\angle$ Cr - O2  - Cr ($^{\rm o}$) & 158.2(3) & 156.50(15) \\
\end{tabular}
\end{ruledtabular}
\end{table}

The high absorption cross section of the element Eu with the natural isotope composition usually prevents high-resolution NPD measurements on Eu substituted samples.   In order to follow the variation of the crystal structure and especially the shifts of the oxygen atoms of  Eu substituted samples, Ce$_{1-x}$Eu$_x$CrO$_3$,  we therefore used the lattice and positional parameters determined from the refinement of the XRPD patterns.
The XRPD patterns of a total of nine phases of Ce$_{1-x}$Eu$_x$CrO$_3$ (0 $\leqslant$ x $\leqslant$ 1) and that of GdCrO$_3$ were measured and Rietveld refinements were performed. Fig. \ref{Survey}(a - c) displays three characteristic diffraction patterns covering the entire range of  $x$ compared  with the results of the  refinements and the difference between observed and calculated patterns.  In addition to the lattice parameters and the atom positions shown in the following figures (cf. Fig. \ref{Lattpar} - \ref{Apical}),  background and  displacement parameters have also been refined which will not be discussed any further.

Fig. \ref{Lattpar} summarizes the lattice parameters and the cell volume of the phases Ce$_{1-x}$Eu$_x$CrO$_3$ versus the Eu content.  The cell volume decrease linearly with a magnitude  which is in good agreement with previous findings.\cite{Gonjal2013}
Whereas  $a$ and $c$ follow a Vegard's-law  with a linear dependence on $x$, $b$ exhibits  non-linear behavior with a bending away from a linear relationship towards large $x$. Remarkable is also the crossing of the lattice parameters $a$ and $c$ at $x \sim$ 0.04 (see inset Figure \ref{Lattpar}) which  similarly has been reported by Prado-Gonjal \textit{et al.} to lie between LaCrO$_3$ and PrCrO$_3$.\cite{Gonjal2013}

\begin{figure}[htp]
	\centering
	\includegraphics[width=8.5cm]{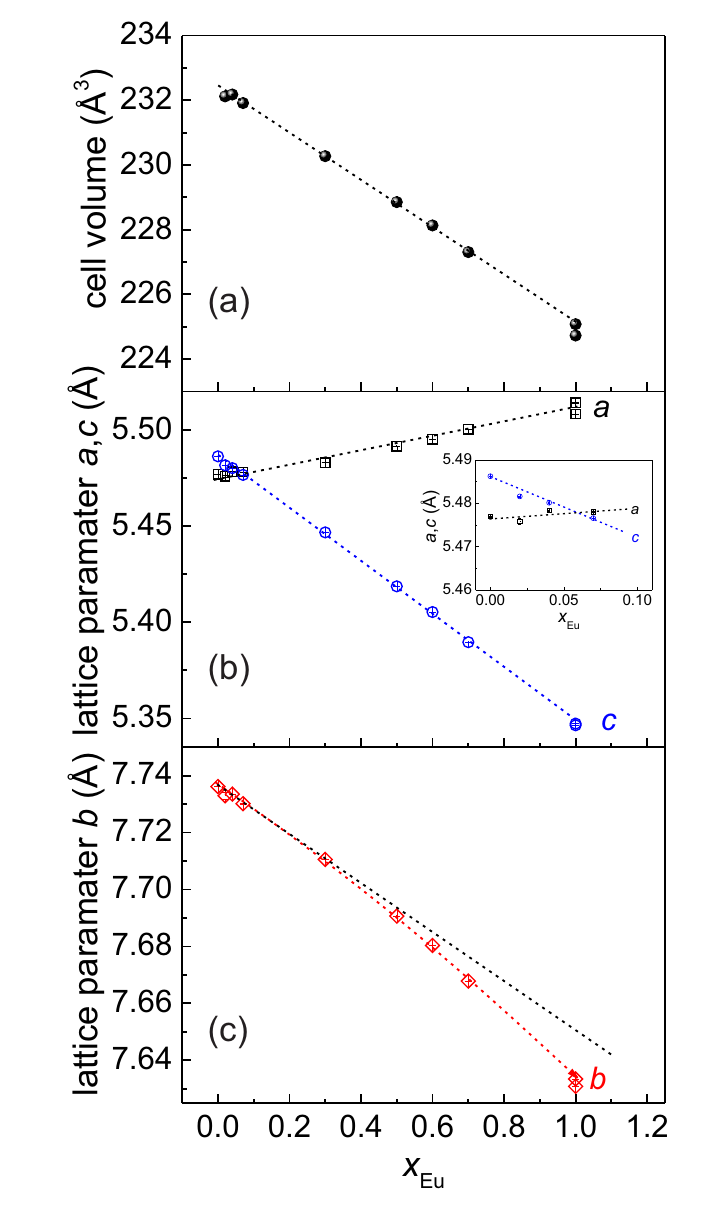}
	\caption{(color online) (a) Cell volume and  lattice parameters $a$, $b$ and $c$ versus the Eu content $x$ for the phases Ce$_{1-x}$Eu$_x$CrO$_3$ ($Pnma$ setting). The inset shows the crossing of the lattice parameters $a$ and $c$ for small $x$. The colored dashed lines are guides to the eye. The black dashed line in (c) emphasizes the non-Vegard's behavior of the lattice parameter $c$.}
	\label{Lattpar}
\end{figure}

The  cell distortion factor, $d$,  introduced by Sasaki \textit{et al.} to measure the deviation of distorted perovskite-type structures from cubic symmetry is estimated according to\cite{Sasaki1983}

\begin{equation}
d = 10^6 \times (a/\sqrt{2}-a_p)^2+(b/2-a_p)^2+(c/\sqrt{2}-a_p)^2)/a_p^2/3,
\label{DistEq}
\end{equation}

where the average 'cubic' lattice parameter $a_p$ is given by

\begin{equation}
a_p =  (a/\sqrt{2}+b/2+c/\sqrt{2})/3.
\end{equation}

$d$ shows a maximum distortion of $\sim$ 150 ppm for EuCrO$_3$ (cf. Fig. \ref{Figure9}), whereas for CeCrO$_3$ $d$ almost vanishes. Compared to the high pressure data by Venkata \textit{et al.}\cite{Venkata2013} a cell distortion factor of 150 ppm is comparable to  distortion factors obtained for example in EuCrO$_3$ for pressures of $\sim$ 10 GPa. Remarkable is the negative slope of the distortion factor, for small, values of $x \leq$0.04 below the intersection of the lattice parameters $a$ and $c$ (cf. inset Fig. \ref{Figure9}) paralleling the crossing of the lattice parameters $a$ and $c$ in this concentration range.

\begin{figure}[htp]
	\centering
	\includegraphics[width=8.5cm]{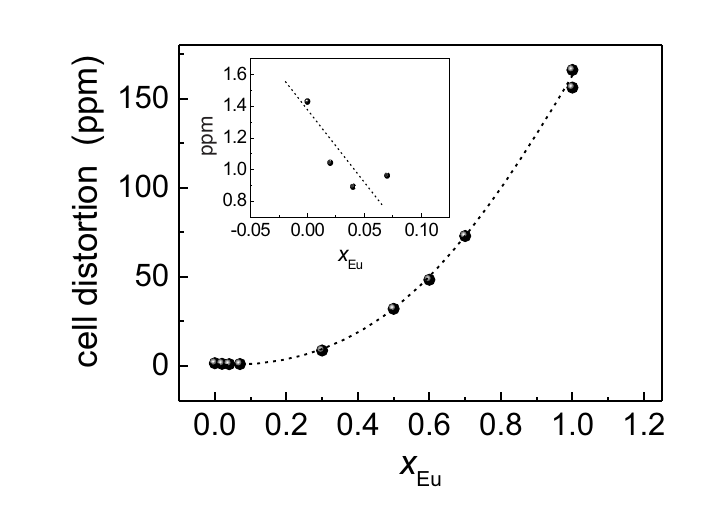}
	\caption{(color online) Cell distortion factor for Ce$_{1-x}$Eu$_x$CrO$_3$ calculated according to Eq. (\ref{DistEq}).}
	\label{Figure9}
\end{figure}

The gradual replacement of the larger Ce$^{3+}$ cations by the smaller Eu$^{3+}$ cations leads to systematic shifts of the atoms in the crystal structures of the phases Ce$_{1-x}$Eu$_x$CrO$_3$  with about the same magnitude for all atoms involved (R, O1, and O2).
Figure \ref{Figure6}  summarizes our  results for the R and the O1 positional parameters.
The $z$ and $x$ coordinates of the Ce/Eu atoms are linearly related. Both are slightly decreasing with a growing Eu content.
In  the inset to Fig. \ref{Figure6}(a), we highlight a projection of the crystal structure of Ce$_{1-x}$Eu$_x$CrO$_3$ along [010] with arrows representing the atom shifts resulting from a 80\% replacement ($x$ = 0.8) of Ce by Eu cations.  The  Ce/Eu atoms  move  in the $a$ - $c$ plane  away from the symmetric (1/2,1/4,1/2) position with the movement along $<$100$>$ being somewhat more pronounced.
In contrast to  the R atoms, the tilting of the apical oxygen atoms O1 away from $<$010$>$ with an increasing Eu content is more complex: As revealed by a plot of the coordinate $z_{\rm O1}$ versus $x_{\rm O1}$ the movement of the O1 atoms  splits into two separate branches with a significantly different dependence on $x$.  For a small Eu content ($x <$ 0.04) $z_{\rm O1}$ is almost independent of $x_{\rm O1}$, whereas for ($x >$ 0.04) one observes a linear relationship of $z_{\rm O1}$ versus $x_{\rm O1}$ which reflects a correlated tilting of the CrO$_6$ octahedra into the $a$ - $c$ plane away from a stretched configuration.

\begin{figure}[htp]
	\centering
	\includegraphics[width=8.5cm]{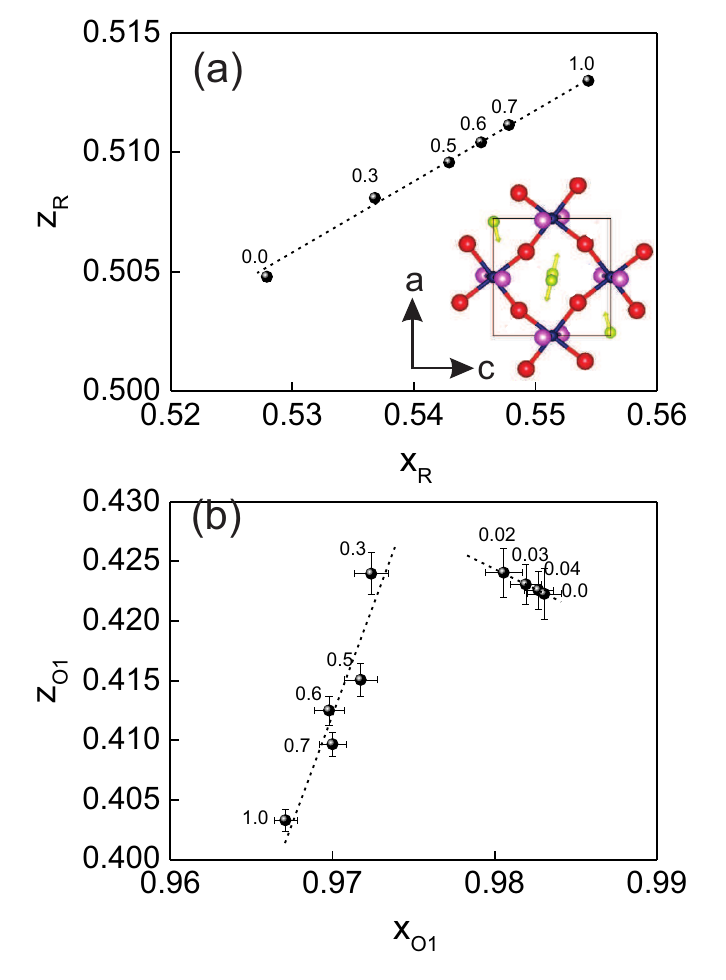}
	\caption{(color online)   (a) Atom positional parameters  of the R atoms versus the Eu content ($x$) for the Ce$_{1-x}$Eu$_x$CrO$_3$ phases. Error bars are of the size of points. The inset  highlights a projection of the crystal structure along [010] with arrows representing the
shifts of the R atoms in the crystal structure  initiated by the replacement of 80\% of the Ce  by Eu atoms, i.e. by comparing phases with $x$ = 0 and $x$ = 0.8.  The length of the arrows is proportional to the atom shifts. (b) Atom positional parameters $z_{\rm O1}$, $x_{\rm O1}$ of the apical oxygen atoms O1 in the CrO$_6$ octahedra. The (black) dashed lines are guides to the eye. The numerals give the Eu content $x$.}
\label{Figure6}
\end{figure}

The apical oxygen atoms O1 move only slightly away from  a orthogonal orientation with respect to the Cr - O2 equatorial plane in the CrO$_6$ octahedra. Figure \ref{Apical} shows a maximum "skewness" of less than 2$^{\rm o}$ with an opposite movement of $\angle$ (O22 - Cr - O1) and  $\angle$ (O21 - Cr - O1) around an average value of $\sim$90 deg.
This tilting of the CrO$_6$ octahedra leads to an increased torsion of neighboring CrO$_6$ octahedra with respect to each other (cf. Figure \ref{Figure7}(b)). The torsion angle
O1 - Cr - Cr - O1 exhibits a linear dependence on the Eu content and increases markedly by about 22\%. The torsion angle is linearly correlated with  the Cr - O2 - Cr  bonding angle (cf. Figure \ref{Figure7}(c)). The latter
characterizes the opposite movement of the O2 equatorial oxygen atoms outward and inward away from a 180$^{\rm o}$ Cr - O - Cr bond as  displayed in Fig. \ref{Figure7}(a).

\begin{figure}[htp]
	\centering
	\includegraphics[width=8.5cm]{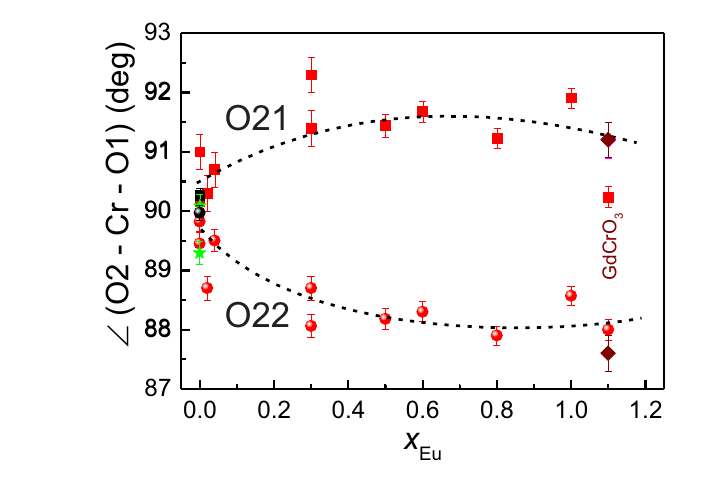}
	\caption{(color online)  O2 - Cr - O1 angles  derived from our neutron diffraction data (black circles and squares) and the x-ray diffraction (red circles and squares)  measurements. Dashed lines are guides to the eye.  Literature data are also plotted: (green) asterisks CeCrO$_3$\cite{Shukla2009}; (brown) diamond GdCrO$_3$\cite{PradoCif}.}
	\label{Apical}
\end{figure}

\begin{figure}[htp]
	\centering
	\includegraphics[width=9cm]{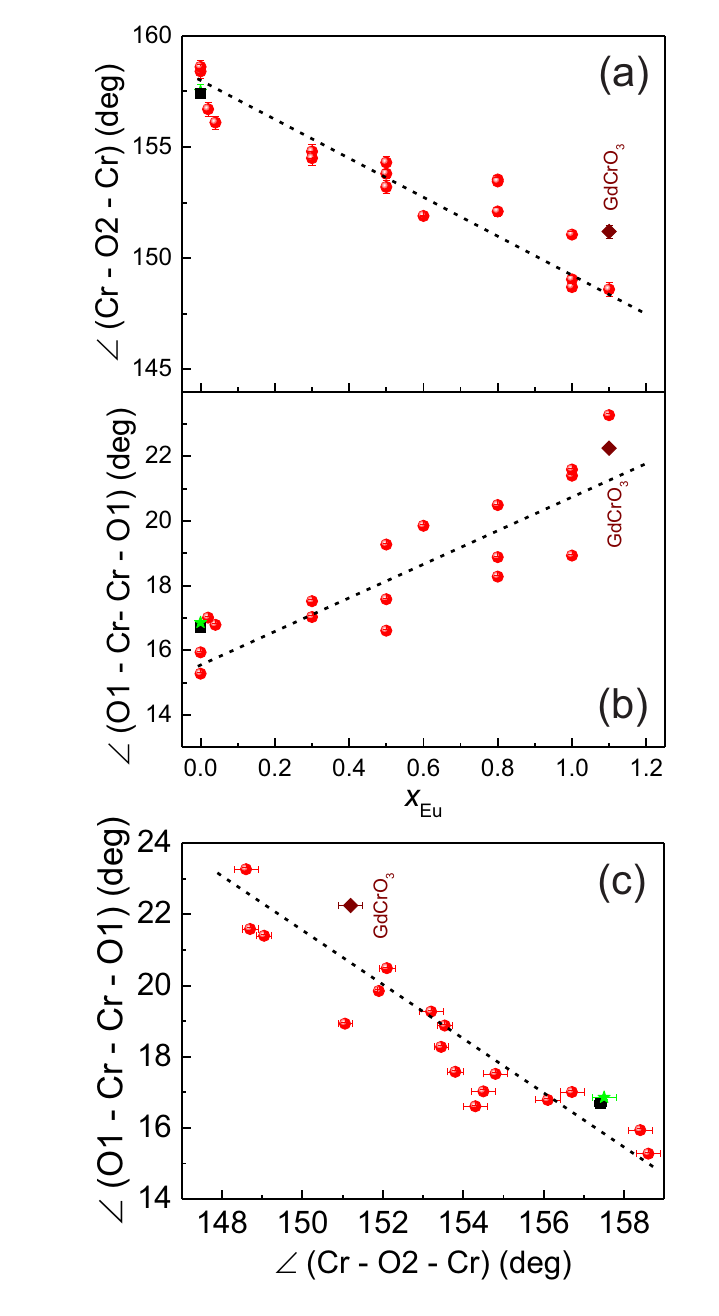}
	\caption{(color online) (a) Cr - O2 - Cr  bonding angles versus Eu content. Plotted is the bonding angle away from a line connecting two Cr atoms.  (b) Torsion angle O1 - Cr - Cr - O1. (c) Torsion angle of (a) versus bonding angle shown in (b). Bonding angles calculated from literature data  are also plotted:  (green) asterisk: CeCrO$_3$\cite{Shukla2009}; (brown) diamond: GdCrO$_3$\cite{PradoCif}.}
	\label{Figure7}
\end{figure}

\begin{figure}[htp]
	\centering
	\includegraphics[width=8.5cm]{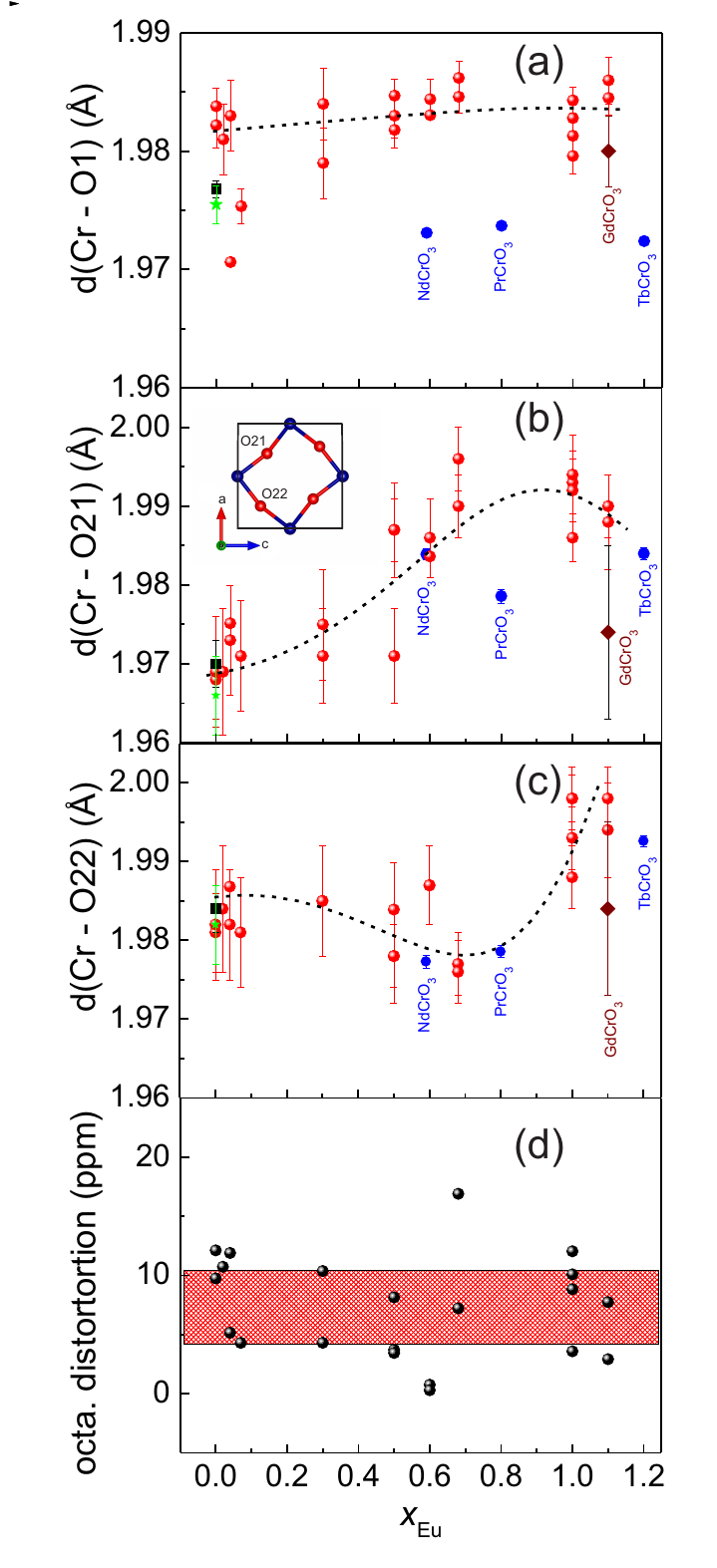}
	\caption{(color online) (a - c) (black) square Cr-to-oxygen interatomic distances derived from our neutron powder diffraction data and the x-ray diffraction (red) circles patterns. Dashed lines are guides to the eye. The upper left inset in (b) gives the assignment of the O21 and O22 atoms. Octahedral distortion factor according to Eq. (\ref{OctaDist}). Literature data are also plotted: (blue) circles RCrO$_3$ (R = Nd, Pr, Tb)\cite{Zhou2010}; (green) asterisk CeCrO$_3$\cite{Shukla2009}; (brown) diamond GdCrO$_3$\cite{PradoCif}. (d) The octahedral distortion factor calculated from the distances in (a) - (c) according to Eq. (\ref{OctaDist}). The dashed area indicates an interval of 3$\sigma$ around an mean value of 7.5$ \pm$ 0.9.}
	\label{VolandDist}
\end{figure}

Figure \ref{VolandDist} summarizes the Cr - O distances in the CrO$_6$ octahedra.
The Cr - O2 distances (Cr-O21 and Cr - O22, i.e. the distance to the O atoms which lie essentially in the $a$ - $c$ plane in $Pnma$ setting) exhibit an opposed $x$ dependence with about the same magnitude of the slopes. The Cr - O1 distance to the apical oxygen atoms essentially parallel $b$ undergoes a moderate elongation.  An estimation of the octahedral distortion factor which we defined in analogy of the cell distortion factor according to

\begin{equation}
d_{\rm oct} = \frac{1}{6}\sum_{i=1}^{6} (d_{i{\rm {Cr - O}}}- d_{\rm ave})^2,
\label{OctaDist}
\end{equation}

where the mean distance, $d_{\rm ave}$ is calculated from the six  Cr to oxygen distances in the CrO$_6$ octahedra, and $d_{i{\rm {Cr - O}}}$, using

\begin{equation}
d_{\rm ave} = \frac{1}{6}\sum_{i=1}^{6} d_{i{\rm {Cr - O}}},
\end{equation}

indicates an octahedral distortion between 5 and 10 ppm (cf. Figure \ref{VolandDist}(d)), largely independent of the Eu content.

\subsection{Magnetic Properties}
\begin{figure}[htp]
\includegraphics[width=8.5cm]{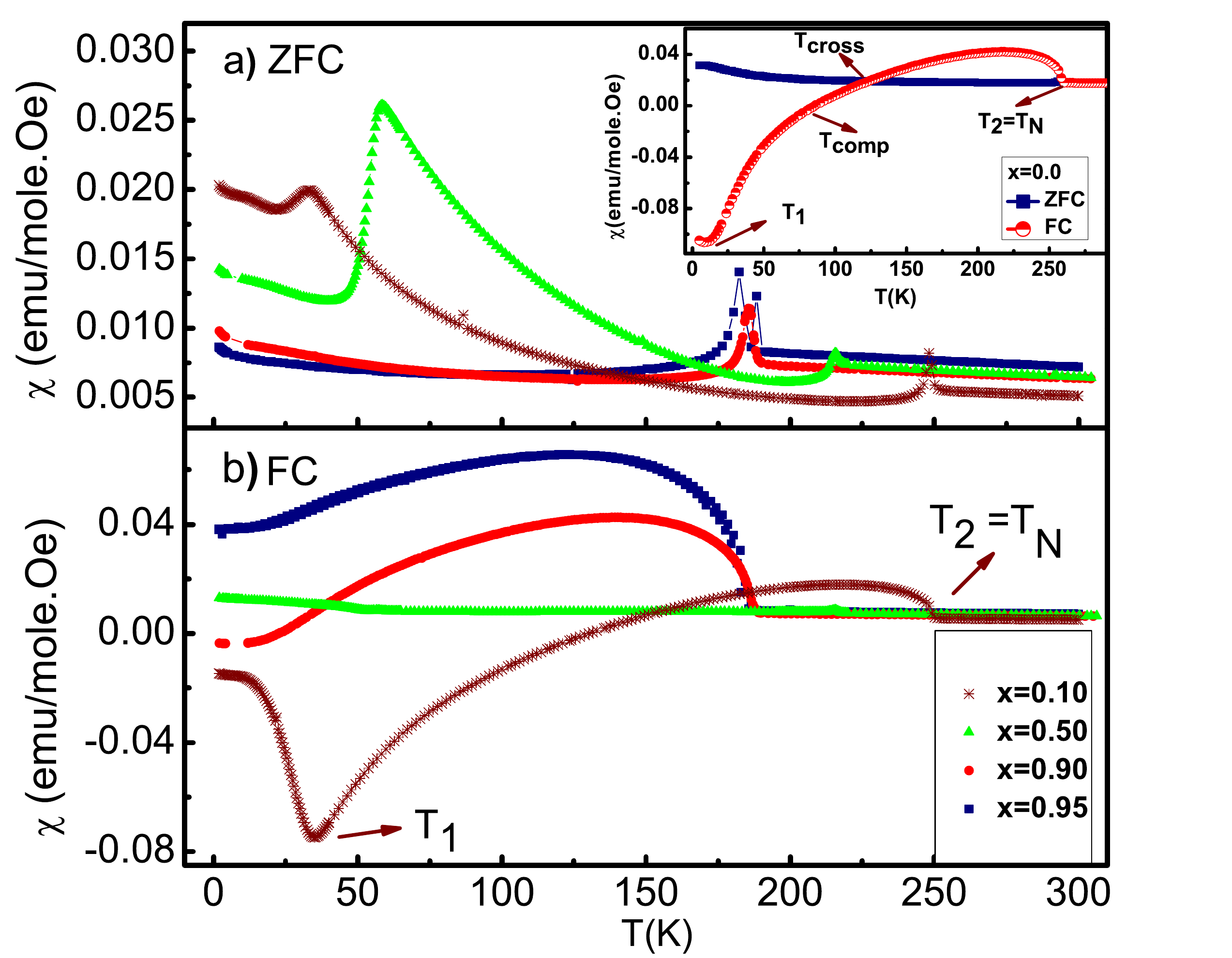}
\caption{(online color) (a) ZFC and (b) FC susceptibilities versus temperature of polycrystalline samples of Ce$_{1-x}$Eu$_x$CrO$_3$ ($x$= 0.1, 0.5, 0.9 and 0.95) under an applied magnetic field of 500 Oe. The inset in (a) shows the temperature dependence of ZFC and FC susceptibilities of CeCrO$_3$.}
\label{SuceptZ(FC)}
\end{figure}

The temperature dependence of the dc magnetizations in an applied magnetic field of $\mu_0H$ = 0.05 T of several polycrystalline samples of Ce$_{1-x}$Eu$_x$CrO$_3$ was measured  in a warming cycle after the samples had been slowly cooled  ($\sim$1.5 K/min) either with zero field (ZFC)  or magnetic field  (FC).
The ZFC and FC data summarized in Figure \ref{SuceptZ(FC)} reveal an antiferromagnetic ordering  of the Cr moments. The N\'{e}el temperature  decreases  from 260 K (CeCrO$_3$) to 178 K (EuCrO$_3$). Small peaks are seen in the ZFC traces whereas the FC magnetizations exhibit a sharp increase  reminiscent of the built-up of spontaneous magnetization in a ferro- or ferrimagnet. These anomalies become more pronounced with increasing Eu content $x$. In addition, the ZFC magnetizations show maxima below $\sim$ 70 K, especially for smaller Eu contents ($x <$ 0.5) and negative magnetization which is attributed to the development of rare-earth magnetic ordering.
The inset in Figure \ref{SuceptZ(FC)}(a) shows, for example, the ZFC (blue solid line) and the FC (red solid line) susceptibilities of CeCrO$_3$ (x = 0.0) which  bifurcate from each other at a splitting temperature of $\sim$258 K. By warming up from 1.8 K to room temperature, {$\chi$}(ZFC)  decreases gradually until it reaches the transition temperature at $T_{\rm N}$ $\sim$ 260 K in good agreement with  previous studies on orthochromites.\cite{Jaiswal,Su,Shukla2009,Cao}  {$\chi$}(ZFC) is positive at all temperatures, whereas in FC measurements the  susceptibility (red line) increases up to the transition temperature of $T_1$ $\sim$12 K and then decreases again to become zero at the compensation temperature, $T_{\rm comp}$. Below $T_{\rm comp}$, the magnetization is negative while with further warming, it changes back to  positive values. Above $T_{\rm comp}$, the magnetization increases to attain a maximum at $T_b$ $\sim$220 K and then decreases gradually to reach the transition temperature  $T_{\rm N}$ $\sim$260 K. The $T_{1}$ anomaly may be ascribed to the magnetic ordering of cerium ions which similarly has been seen as the spin reorientation in different members of RCrO$_3$ compounds \cite{Shukla2009, Yoshii, Tiwari}. Interestingly, Ce ordering appears to have no immediate impact on the Cr magnetic structure as  the intensity of the (110)/(011) magnetic Bragg reflection follows an standard order parameter temperature dependence (see discussion and Figure \ref{Neutron} below).

\subsection{Heat Capacities}
\begin{figure}[htp]
	\centering
	\includegraphics[width=8.5cm]{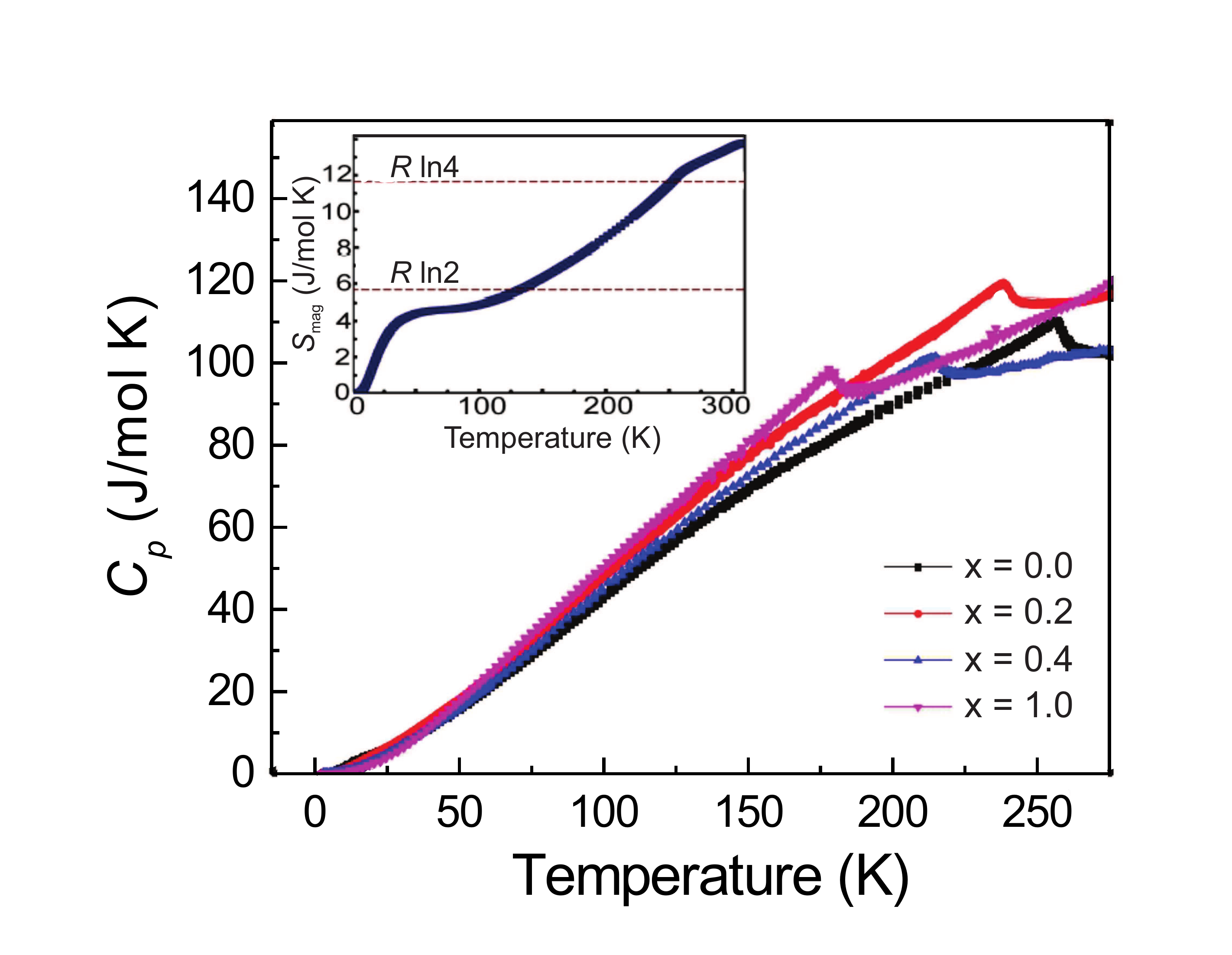}
	\caption{(color online) Temperature dependence of total heat capacities of Ce$_{1-x}$Eu$_x$CrO$_3$ ($x$ = 0.0, 0.2, 0.5, 1.0). The inset shows the temperature dependence of magnetic entropy of CeCrO$_3$ ($x$ = 0.0).}
	\label{HC-All}
\end{figure}

The heat capacities of most rare-earth orthochromites, RCrO$_3$ (R = La. Pr, Nd, Sm. Gd, Dy, Ho, Er and Y) have been reported by Satoh \textit{et al.} \cite{Satoh1997}, Bartolom\'{e} \textit{et al.}  and Su \textit{et al.}.\cite{Bartolome2000, Su2010, Su2011}
In this work, we measured  the heat capacities on Ce$_{1-x}$Eu$_x$CrO$_3$ phases down to liquid helium temperature at zero magnetic field and studied the anomalies associated with the long-range magnetic ordering of Cr$^{3+}$ ions.
Figure \ref{HC-All}  displays the temperature dependence of the heat capacity of the Ce$_{1-x}$Eu$_x$CrO$_3$  from 1.8 K to 300 K in zero magnetic field where $x$ = 0.0, 0.2, 0.4, 1.0.
The long-range ordering of the  Cr$^{3+}$ moments indicated by the  $\lambda$-shaped anomaly shifts   from 256 K for $x$ = 0.0 to 175 K for $x$ = 1.0.
Especially for small Eu contents, the
heat capacities at low temperatures revealed additional magnetic contributions. These can be clearly separated from the total heat capacities when the heat capacities of the lattice were  subtracted. The lattice heat capacities were estimated from the data of the non-magnetic isotypic polycrystalline sample\cite{MaryamPhD,MaryamPRB} of LaGaO$_3$ by
scaling the temperature according to the different molar masses of RCrO$_3$ and LaGaO$_3$ following  the procedure proposed by Bouvier \textit{et al.}.\cite{Bouvier}

The magnetic entropy of CeCrO$_3$ gained by integrating $C_{\rm mag}/T$ vs. $T$, is plotted in  the inset in Figure \ref{HC-All}.
At low temperatures, the entropy amounts to $\sim R ln\,2$ indicating a doublet crystal field ground state of Ce$^{3+}$. Towards higher temperatures, the entropy captures contributions from excited crystal field states of Ce$^{3+}$ and contributions of Cr$^{3+}$ moment ordering. At the N\'{e}el temperature, the entropy  reaches a value of  $\sim R ln\,4$.

\subsection{Magnetic Neutron Scattering}
\begin{figure}[htp]
\centering
\includegraphics[width=9cm,height=8.5cm]{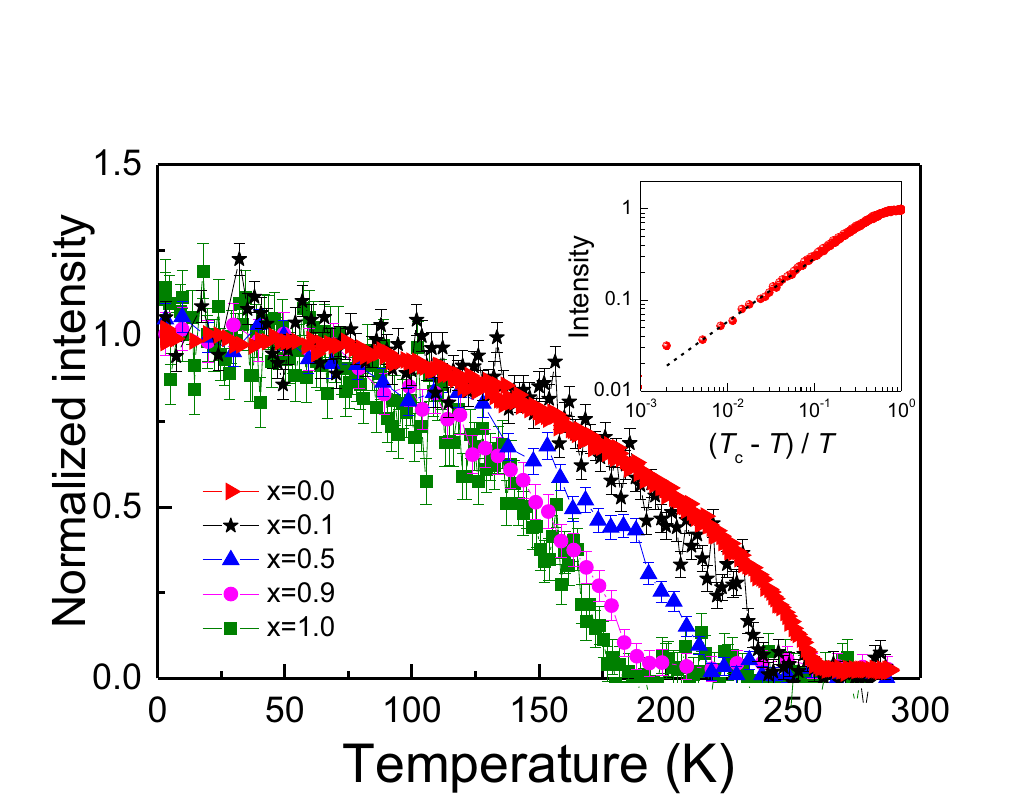}
   \caption{(online color) Integrated intensity of the (110)/(011) magnetic Bragg reflection at 2$\theta \sim$ 30.9 $^{\rm o}$ vs temperature for various samples of Ce$_{1-x}$Eu$_x$CrO$_3$ as indicated. The inset shows the intensity versus the reduced temperature, $t$ =($T_N$ - $T$)/$T_N$, where the critical temperature, $T_N$,  for CeCrO$_3$ amounts to 258.1(1) K. The dashed line represents a critical power law $\propto$ $t^{2 \beta}$ with a critical exponent $\beta$ = 0.346(5). }
 \label{Neutron}
\end{figure}

In order to determine the exact values of $T_{\rm N}$, temperature dependent low-resolution high-intensity neutron powder diffraction patterns was carried out on several samples of Ce$_{1-x}$Eu$_x$CrO$_3$.
Below the N\'{e}el temperature, all patterns exhibited a pronounced magnetic Bragg reflection near  2$\theta \sim$ 30.9 $^{\rm o}$ which can be indexed as (110)/(011). Additional, apparent magnetic scattering intensity was noticed for the (211)/(031) similar to what has been reported for CeCrO$_3$ by Shukla \textit{et. al} before.\cite{Shukla2009} The integrated intensities of the (110)/(011) magnetic Bragg reflection are compiled in Figure \ref{Neutron}. The integrated intensities follow a typical temperature dependence with saturation at low temperatures.
Near to $T_{\rm N}$ the temperature dependence of the intensities for CeCrO$_3$ was  fitted to a critical power law (($T_{\rm N}-T$)/$T_{\rm N}$)$^{2\beta}$ as shown in the inset to Figure \ref{Neutron} resulting in critical exponents, $\beta$, close to 0.34 in agrement with values expected for typical 3dim universality classes. None of the temperature traces for the different compositions reveals any signatures for re-orientation of the Cr moments  at low temperatures associated to the anomalies observed in the magnetic susceptibilities (cf. Figure ref{SuceptZ(FC)} indicating weak exchange coupling between the Ce moments. These rather orient readily in a stronger exchange fields of the Cr moments without a feed back.

\section{Discussion and Summary }

\begin{figure}[htp]
	\centering
	\includegraphics[width=8.5cm]{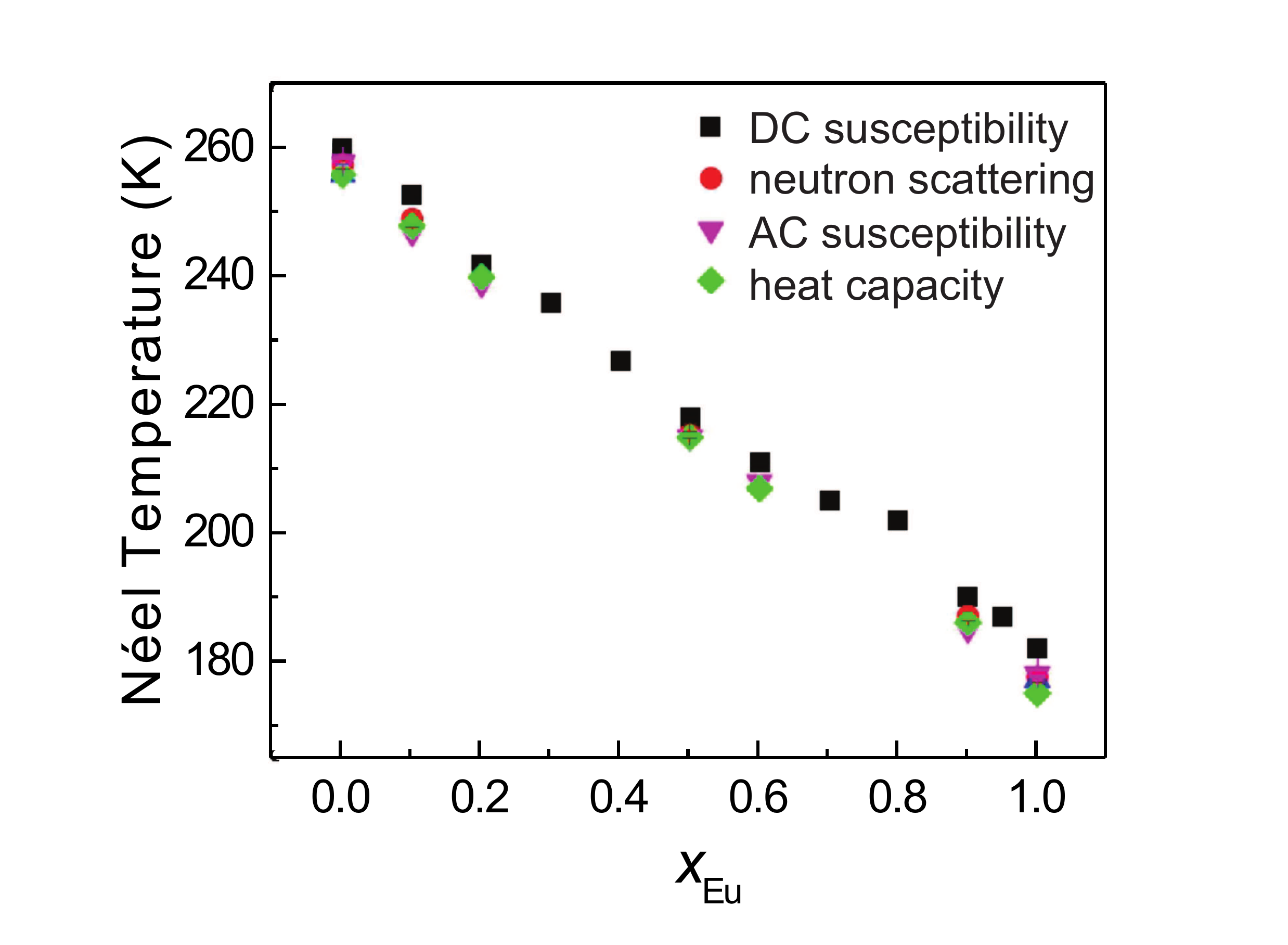}
	\caption{(color online) N\'eel transition temperature ($T_{\rm N}$) versus europium concentration, $x$, in Ce$_{1-x}$Eu$_x$CrO$_3$, obtained from the different techniques used in this work.}
	\label{TN}
\end{figure}

We have synthesized high-purity ceramic samples of Ce$_{1-x}$Eu${_x}$O$_3$ where $x$ varied from 0 to 1. Extensive studies of the structural parameters by neutron and x-ray powder diffraction measurements reveal the dependence of antiferromagnetic ordering on details of the bonding angles and the lattice size. The  effect of the  cell decrease on the magnetic properties induced by substituting Eu for Ce  appears to be  rather indirect via a systematic change of Cr - O - Cr bonding and torsion angles which become  necessary in order to accommodate the rather rigid CrO$_6$ octahedra in the shrinking lattice.
Substituting Eu for Ce in Ce orthochromites of composition Ce$_{1-x}$Eu$_x$CrO$_3$ not only rapidly decreases the N\'{e}el temperature of the antiferromagnetic ordering of the Cr$^{3+}$ moments, but also gradually changes the  thermal hysteresis behavior observed in the dc magnetization. In fact, samples with higher Eu content  show a ZFC/FC splitting  with a small peak in the FC susceptibility whereas the FC magnetization is reminiscent of the development of spontaneous magnetization in a ferromagnet. However, the saturation magnetization is far too low to account for  the expected saturation magnetization of  $\sim$3$\mu_{\rm B}$ expected for Cr$^{3+}$ with the spin $S$=3/2 moment. This finding rather indicates  canted antiferromagnets as found e.g. for EuCrO$_3$.\cite{MaryamPRB} Using high-resolution x-ray diffraction measurements with sufficient intensity up to $d$-values of $\sim$ 0.4 \AA\ allowed us to reliably refine oxygen atom positions.
From our structural investigation we concluded:
(a) The cell volume decreases by $\sim$2.5\%  from CeCrO$_3$ to EuCrO$_3$ implying an overall increased density of Cr cations per unit volume for increasing Eu content.
(b) Even though the cell volume reduction implies an overall increased density of Cr cations per unit volume, the N\'eel  temperature decreases linearly with the Eu content. Figure \ref{TN} showing the N\'eel transition temperature of Ce$_{1-x}$Eu$_x$CrO$_3$ versus Eu content summarizes the results of   all magnetic characterization measurements performed  on the various phases studied in this work.

The N\'{e}el temperatures decrease linearly with the Eu content  following a Vegard's-type  dependence on the Eu concentration  according to

 \begin{equation*}
T_{\rm N} = 260 K - x \times 80 K.
\end{equation*}

An analogous decrease of $T_{\rm N}$ induced by a decreasing cell volume   across the whole  series of the rare earth orthochromites has been observed before.\cite{Zhou2010,Gonjal2013} From the light to  the heavier rare-earth elements $T_{\rm N}$  of RCrO$_3$  decreases by 60\%. In terms of lanthanide contraction, the solid-solutions investigated in this work cover approximately half of the rare-earth series. By adjusting the Ce/Eu ratio, they allow to divide the interval between CeCrO$_3$ and EuCrO$_3$ into several small intervals which enables us  to relate $T_{\rm N}$ with  minute structural changes. Substituting Ce by the smaller Eu cations exerts chemical pressure.
Using first-principles calculations Zhao \textit{et al.} have estimated the effects of  chemical pressure external versus hydrostatic pressure on the N\'{e}el temperature of rare-earth orthochromites.\cite{Zhao2013} They found that hydrostatic pressure essentially modifies bond distances and in turn leads to an increase of  the N\'{e}el temperature whereas chemical pressure exerted by substituting a smaller rare-earth cation reduces the N\'{e}el temperatures by modifying  antipolar displacements, Cr - O - Cr bond angles
and the resulting oxygen octahedral tilts.\cite{Zhao2013}

(c) Despite the decrease of the cell volume, the CrO$_6$ octahedra did neither shrink nor distort significantly. The Cr - O1 distances remain almost constant (cf. Fig. \ref{VolandDist}(a)) and one does not expect a marked alteration of the energies of the $t_{2g}$ orbitals. The Cr - O2 distances  (cf. Fig. \ref{VolandDist}(b) and (c)) exhibit a slight opposite dependence on the Eu content, however, their mean value exhibits only a very small, approximately linear increase by less than a percent. Accordingly, the "skewness" of the octahedra remains below $\sim$2$^{\rm o}$  (cf. Fig. \ref{Apical}).

(d),  Most striking, however, are the changes of the oxygen positions in the cell and the Cr - O - Cr bonding and torsion  angles. The  O1 atoms  move considerably into the direction of the $c$ - axis ($\sim$5 \%, cf. Fig. \ref{Figure6}(a)). Apparently, as a consequence of the decrease of the cell parameters and the cell volume and  in order to keep the shape and size of the CrO$_6$ octahedra essentially unchanged, the Cr - O2 - Cr bonding angles decrease by $\sim$6 \% (cf. Figure \ref{Figure7}(a)). Simultaneously, the torsion angle O1 - Cr - Cr - O1 which measures to what extent neighboring CrO$_6$ octahedra are inclined with respect to each other increases by $\sim$25 \% (cf. Figure \ref{Figure7}(a)).
The importance of bonding angles for the superexchange and the magnetic ordering has been investigated in detail by Zhou \textit{et al.} for the RCrO$_3$ phases with emphasis on the heavy rare-earth elements and some light rare-earth R = La, Pr, Nd.  In our investigation we have divided the range from CeCrO$_3$ to EuCrO$_3$, i.e. starting from a rare-earth element with a rather large ionic radius into several intervals and find analogous relationships between magnetic ordering and bonding and torsion angles. Especially the latter has been identified to be very essential for intrasite hybridization between  Cr \textit{t} and \textit{e} orbitals. Intrasite hybridization lifts the \textit{t} degeneracy and supports ferromagnetic coupling versus antiferromagnetic superexchange between the Cr - Cr moments.\cite{Zhou2010}

Finally we want to briefly comment on the  crossing of the lattice parameters $a$ and $c$ (Figure \ref{Lattpar}) and the  behavior of the positional parameters $z_{\rm O1}$ and $x_{\rm O1}$ of the apical oxygen atoms (Figure \ref{Figure6}) being different  for small Eu levels, $x \leq$ 0.04 from the behavior for $x \geq$ 0.3. The intersection of $a$ and $b$ near $x$ = 0.04 also reflected by the very small cell distortion factor (Figure \ref{Figure9}) pointing to a lattice metrics close to  cubic, seemingly indicating an attractive energy minimum which also pins the O1 atom positional parameters. Alternatively, a relaxation of lattice strain in the vicinity of the cubic metrics by random substitution of a substantially smaller Eu cation may also be taken into consideration to explain this observation.

In summary, our highly resolved x-ray powder and neutron powder diffraction measurements reveal that   in the solid solutions Ce$_{1-x}$Eu$_x$CrO$_3$ the CrO$_6$ octahedra remain essentially rigid whereas Cr - O bonding and torsion angles change markedly by gradually substituting Ce by Eu. Simultaneously, the N\'{e}el temperature decreases from 260 K to 180 K. As the O2 - Cr - O2 bonding angle continuously moves away from  180$^{\rm o}$  the torsion angles, which reflects  to what extent the Cr - O2 - Cr bonds rotate out of the $a$ - $c$ plane when the size of the rare-earth ion is gradually reduced. This appears to constitute another determining parameter for magnetic ordering in the solid-solution phases  Ce$_{1-x}$Eu$_x$CrO$_3$.

$^\dagger$ Present address: Department of Chemistry, University of Calgary, Calgary, Canada.

$^\ddagger$ Present address:Santa Clara University, Physics Department, 500 El Camino Real Santa Clara, CA 95053

\section*{Acknowledgment}

This research was financially supported by Brock University, the Natural Sciences and Engineering Research Council of Canada (NSERC), the Ministry of Research and Innovation (Ontario), and the Canada Foundation for Innovation, Canada.

\end{document}